\documentclass{ws-mpla}
\usepackage[super]{cite}
\usepackage{graphicx}

\usepackage{amssymb}
\usepackage{bm}
\usepackage{slashed}

\begin{document}

\title{Two natural scenarios for dark matter particles coexisting with supersymmetry}

\author{\footnotesize Maxwell Throm, Reagan Thornberry, John Killough, Brian Sun, Gentill Abdulla, \\
Roland E. Allen}

\address{Department of Physics and Astronomy, Texas A\&M University \\
College Station, Texas 77843, USA \\ allen@tamu.edu}

\maketitle

\begin{abstract}
We describe two natural scenarios in which both dark matter WIMPs (weakly interacting massive particles) and a variety of supersymmetric partners should be discovered in the foreseeable future. In the first scenario, the WIMPs are neutralinos, but they are only one component of the dark matter, which is dominantly composed of other relic particles such as axions. (This is the multicomponent model of Baer, Barger, Sengupta, and Tata.) In the second scenario, the WIMPs result from an extended Higgs sector and may be the only dark matter component. In either scenario, both the dark matter WIMP and a plethora of other neutral and charged particles await discovery at many experimental facilities. The new particles in the second scenario have far weaker cross-sections for direct and indirect detection via their gauge interactions, which are either momentum-dependent or second-order. However, as we point out here, they should have much stronger interactions via the Higgs. We estimate that their interactions with fermions will then be comparable to (although not equal to) those of neutralinos with a corresponding Higgs interaction. It follows that these newly proposed dark matter particles should be within reach of emerging and proposed facilities for direct, indirect, and collider-based detection.

\end{abstract}

\section{Introduction}

After decades of intense efforts, neither supersymmetry~\cite{Baer-Tata,Kane,Nath,Baer-Barger1,Baer-Barger2,no-susy}  nor dark matter 
particles~\cite{susy-DM-1996,Silk,Strigari,Rauch,Freese,Roszkowski,Aprile,Peskin,Olive1,Olive2,Baer-Barger3,Bertone}  have been detected. 
One should recall, however, that historically important discoveries typically require patient waits -- 48 years for the Higgs boson, a century for gravitational waves, and almost two centuries for black holes. 
There are still compelling motivations for seeking both of these proposed central features of nature: Alternatives to dark matter have been rendered increasingly implausible by astronomical observations; and without supersymmetry (susy) it is hard to understand the unification of coupling constants at high
energy or why the Higgs boson mass is not enormously increased by radiative corrections.

The pessimism regarding susy is in part due to experimental
limits that now rule out the simplest models (minimal supergravity and
the minimal supersymmetric standard model). But there was never any reason
to believe that simplistic models like these would be quantitatively valid.
They have primarily served to provide valuable guidance for the qualitative
role of susy in various physical phenomena.

Another discouraging development was the finding that natural supersymmetric models (which are consistent with experiment) have
difficulty in predicting the observed relic abundance of dark matter, 
\textit{if it is assumed that the dark matter consists entirely of
supersymmetric partners}. But if this assumption is dropped, as in the
scenarios of the next section, the tension between theory and observation is ameliorated~\cite{Baer-Barger3}.

Regarding dark matter searches, the cross-sections were always
known to be small, since observations demonstrate that these particles
cannot interact through the electromagnetic or strong force. The limits that
have been established are consistent with either of the two scenarios in the
next section. On the other hand, both neutralinos and the new particles discussed here can still lie within 
reach of the direct-detection experiments planned for the
next few years, as well as an upgraded LHC, and possibly the AMS and Fermi
satellite experiments.

\section{Two testable scenarios: neutralinos plus axions, and a new WIMP
candidate with mass $\leq $ 125 GeV}

Recently it has been pointed out that a multicomponent dark matter scenario,
dominated by e.g. axions, but with a significant admixture of neutralinos,
relieves the tension between susy dark matter and the observed dark matter
abundance~\cite{Baer-Barger3}. This suggestion provides motivation for both the many WIMP
searches -- including Xenon, LZ, and SuperCDMS -- and the very different
searches for axions.

An alternative scenario is that the (only or principal) dark matter particle is the
one recently proposed in an extension of the Higgs sector~\cite{DM1,DM2}. To facilitate the
discussion below, in which this particle is compared with the neutralino, we
will call particles of this kind (neutral or charged) ``Higgsons'', and will 
represent them by $H$. (They are then to be distinguished from 
Higgs bosons $h$ and their superpartners, the
Higgsinos $\widetilde{h}$.) 

In Figs. 1-5 we show a few of the most basic
interactions of Higgsons in direct detection experiments (Figs.
1-3), indirect detection following annihilation (Fig. 4), and collider
detection after creation by proton-proton collisions (Fig. 5). The first
three processes follow from the action in Eq. (40) of Ref.~\cite{DM2} : 
\begin{equation*}
S_{H}=\sum\limits_{i}\int d^{4}x\,\left( H^{\,i\,\dag }\left(
x\right) D^{\mu }D_{\mu }H^{i}\left( x\right) -\left( \frac{1}{2}
H^{\,i\,\dag }\left( x\right) \,S^{\mu \nu }F_{\mu \nu }\,H^{i}\left(
x\right) +h.c.\right) \right) \;
\end{equation*}
where $i$ labels the various species of neutral plus charged Higgson fields.
\begin{figure}
\centering
\includegraphics[width=0.4\textwidth]{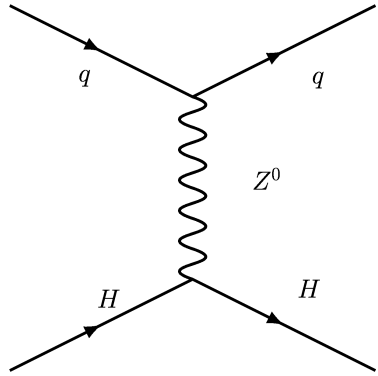} 
\label{fig1}
\caption{Direct detection via $Z^0$ exchange with first-order momentum-dependent vertex.}
\end{figure}
\begin{figure}
\centering
\includegraphics[width=0.4\textwidth]{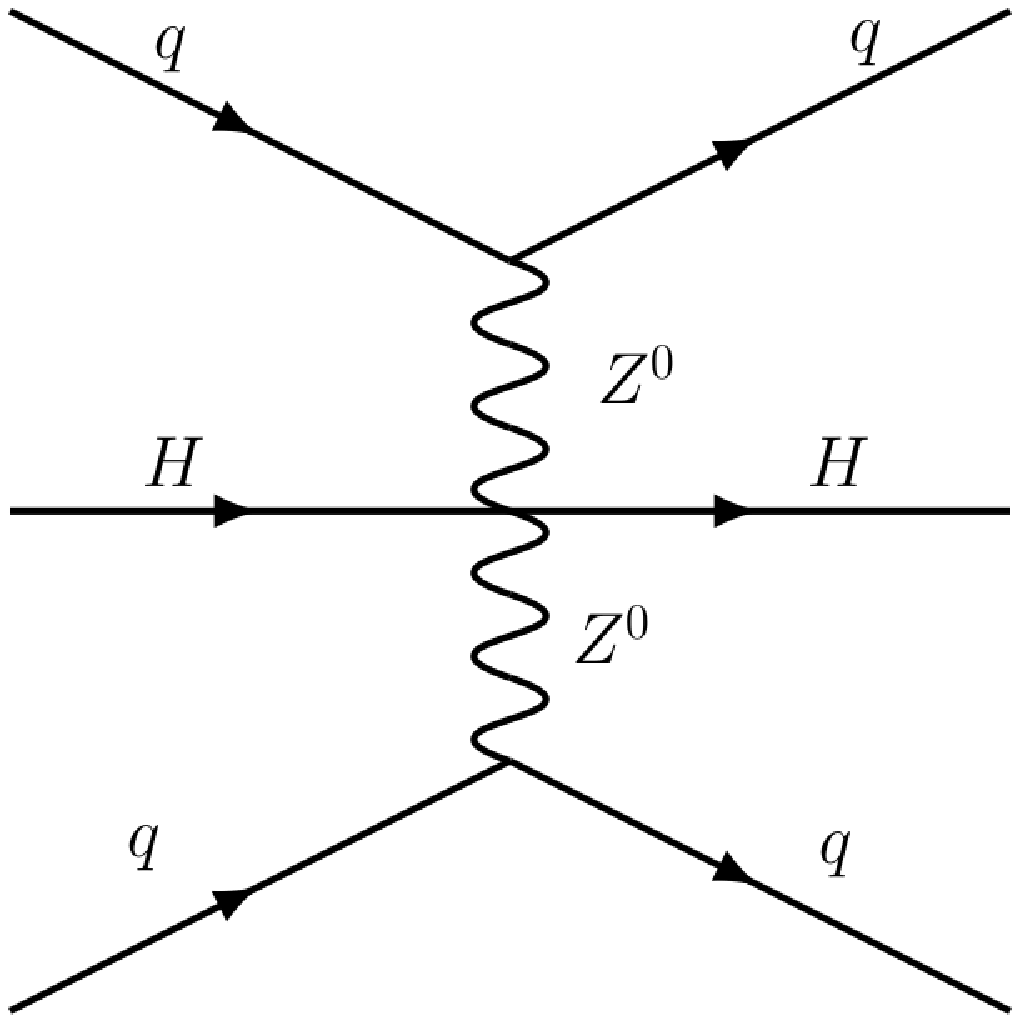} 
\includegraphics[width=0.4\textwidth]{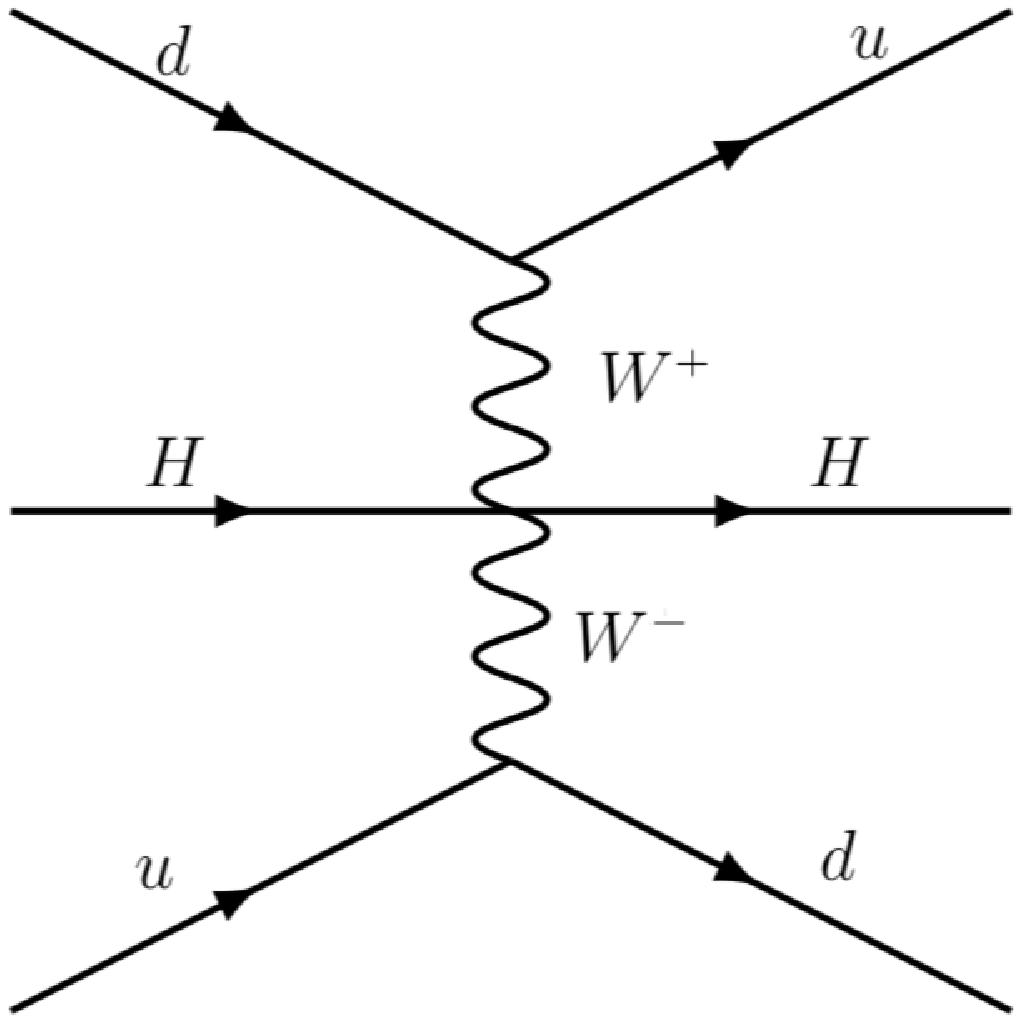} 
\label{fig2}
\caption{Left: Direct detection via double $Z$ exchange with second-order vertex. Right: Direct detection via double $W$ exchange with second-order vertex}
\end{figure}
\begin{figure}
\centering
\includegraphics[width=0.4\textwidth]{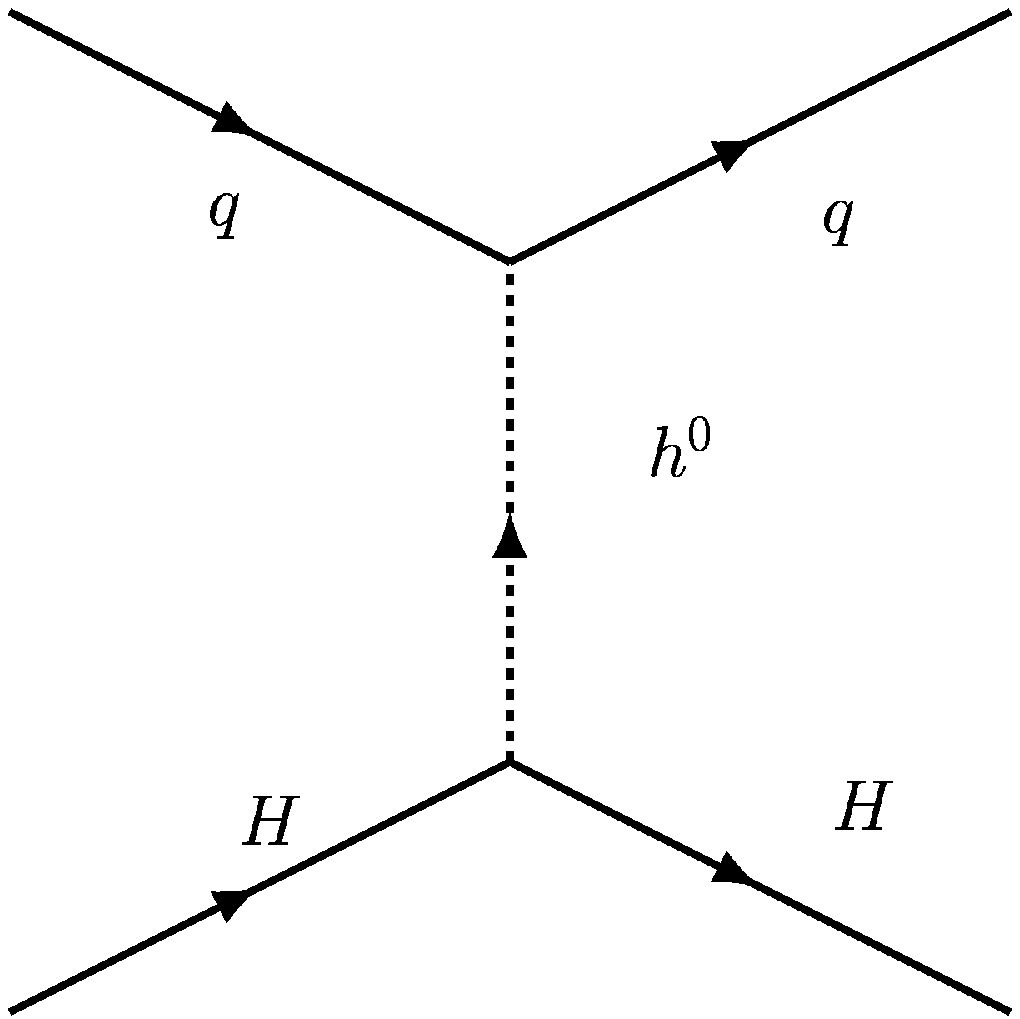} 
\includegraphics[width=0.4\textwidth]{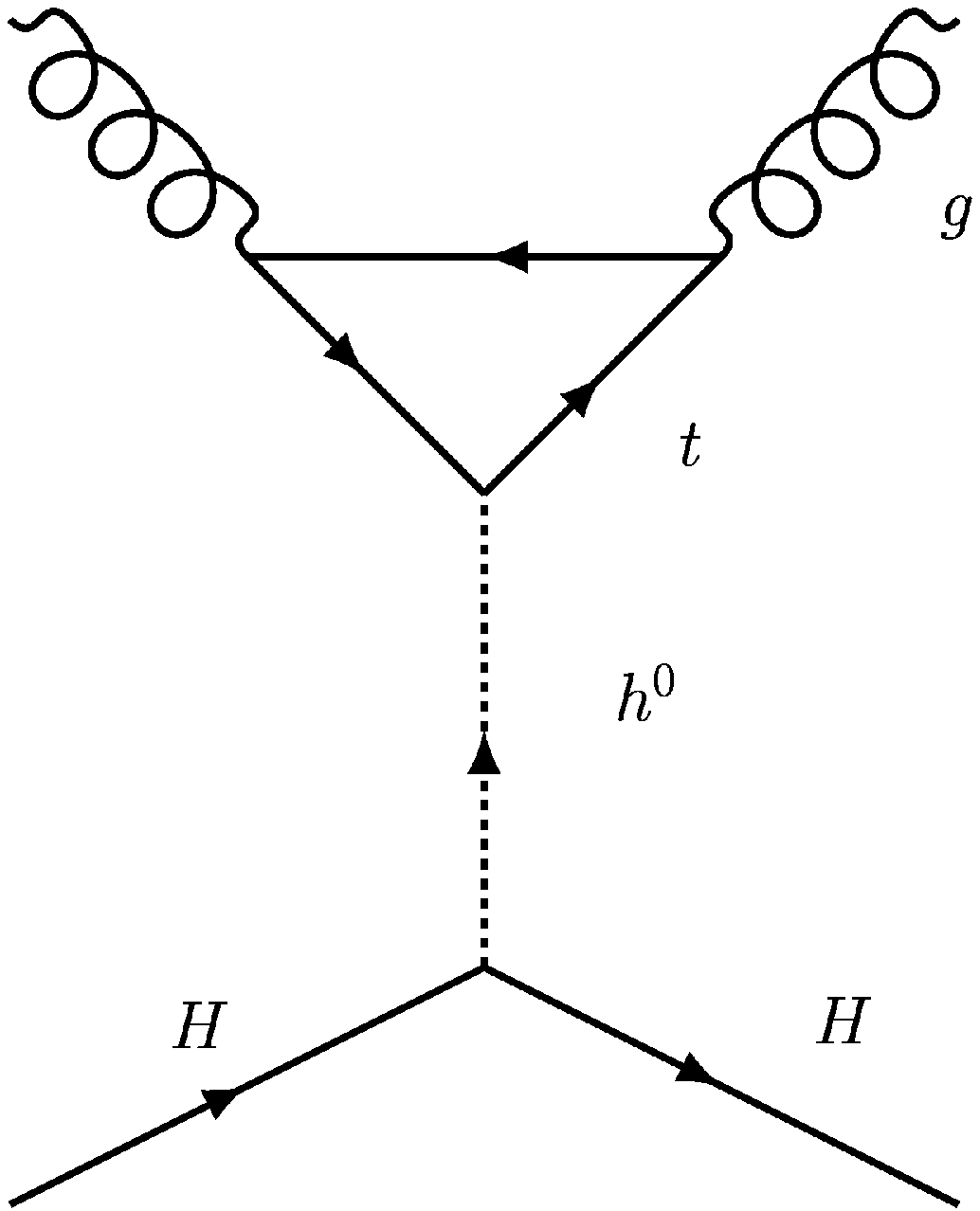} 
\label{fig3}
\caption{Left: Direct detection via $h^0$ exchange with, e.g., strange quark. Right: Direct detection via $h^0$ exchange with top quark triangle coupled to gluons}
\end{figure}
\begin{figure}
\centering
\includegraphics[width=0.49\textwidth]{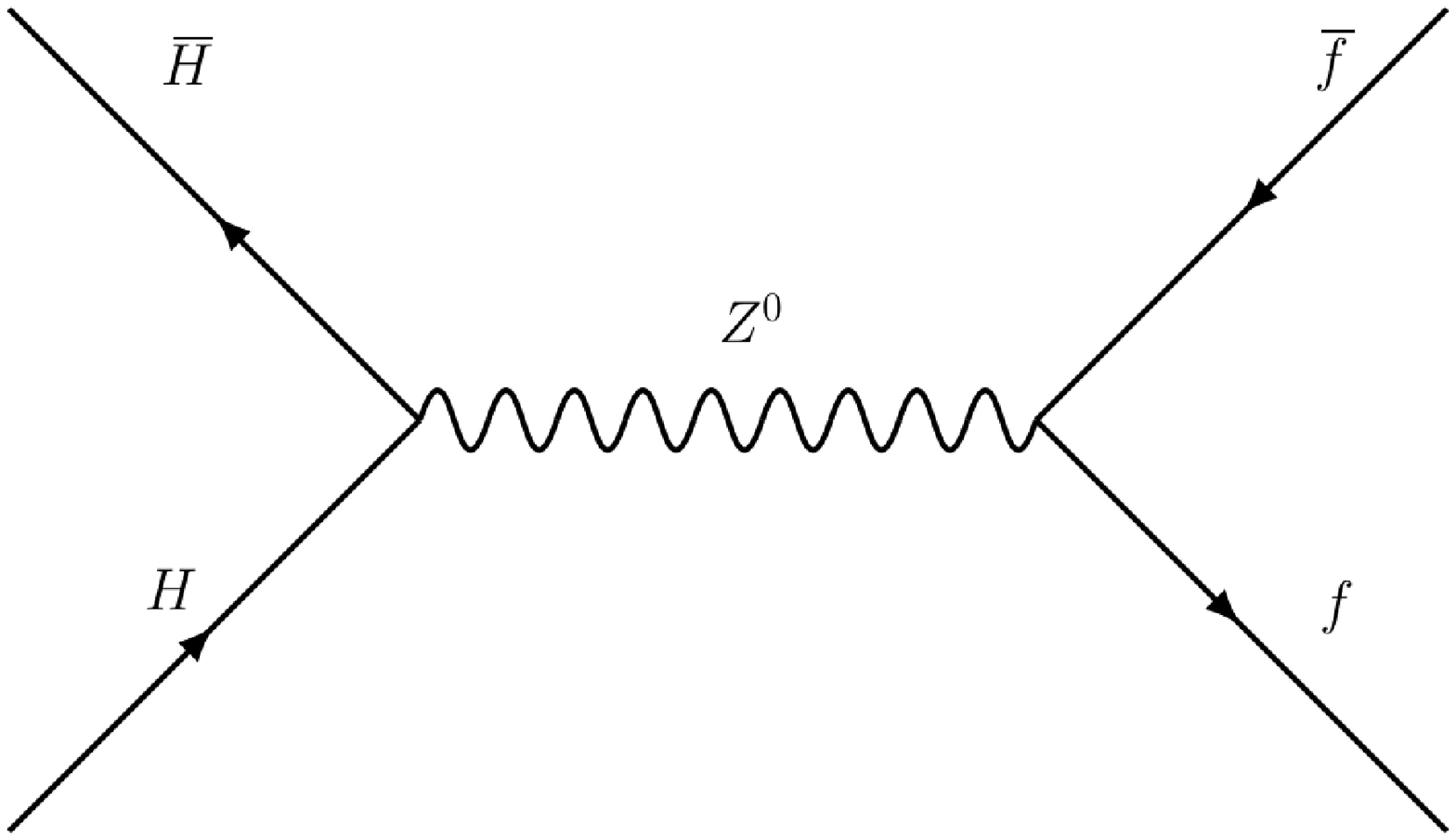} 
\includegraphics[width=0.49\textwidth]{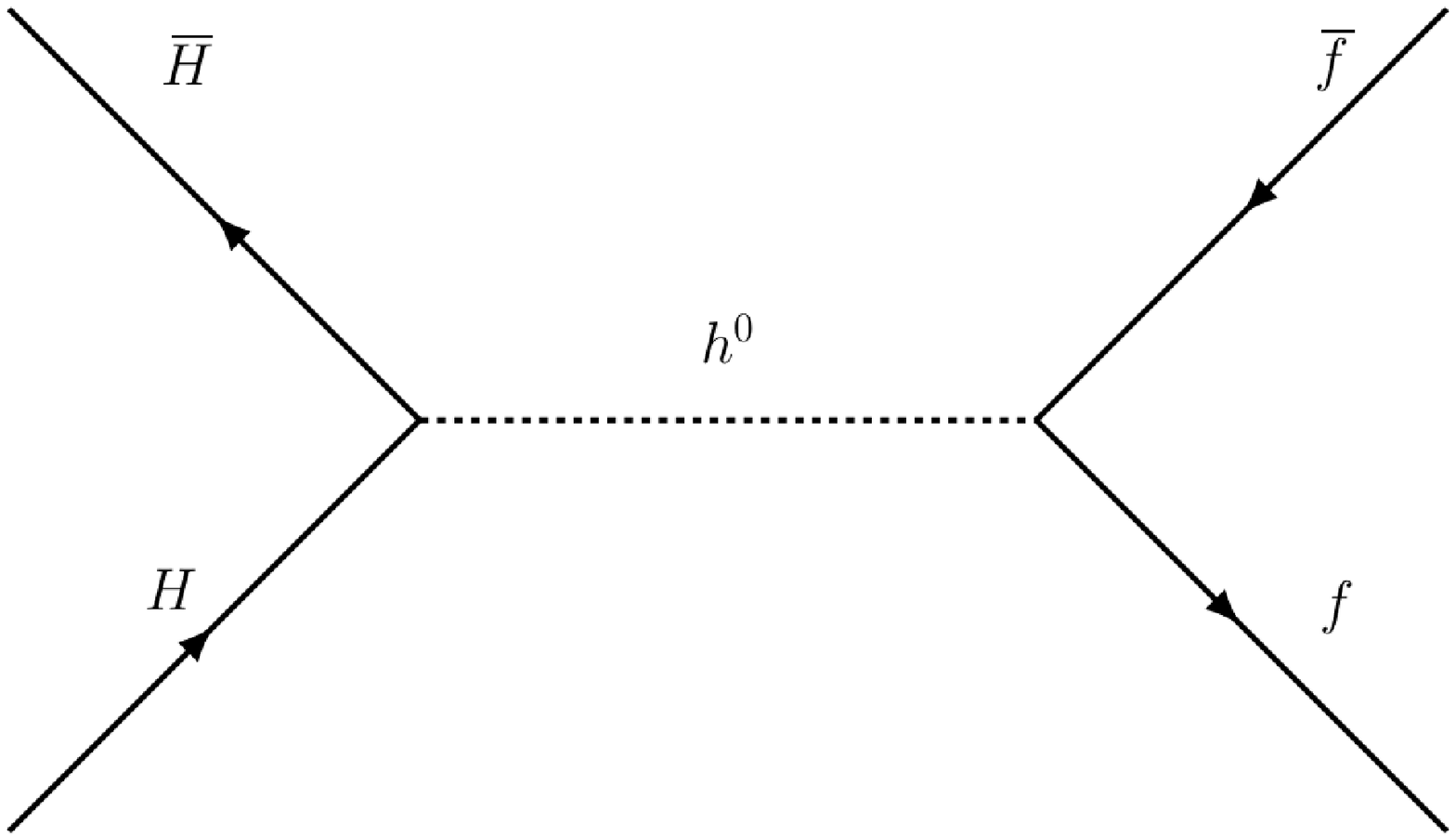} 
\label{fig4}
\caption{Left: Indirect detection via $Z^0$. Right: Indirect detection via $h^0$.}
\end{figure}
\begin{figure}
\centering
\includegraphics[width=0.49\textwidth]{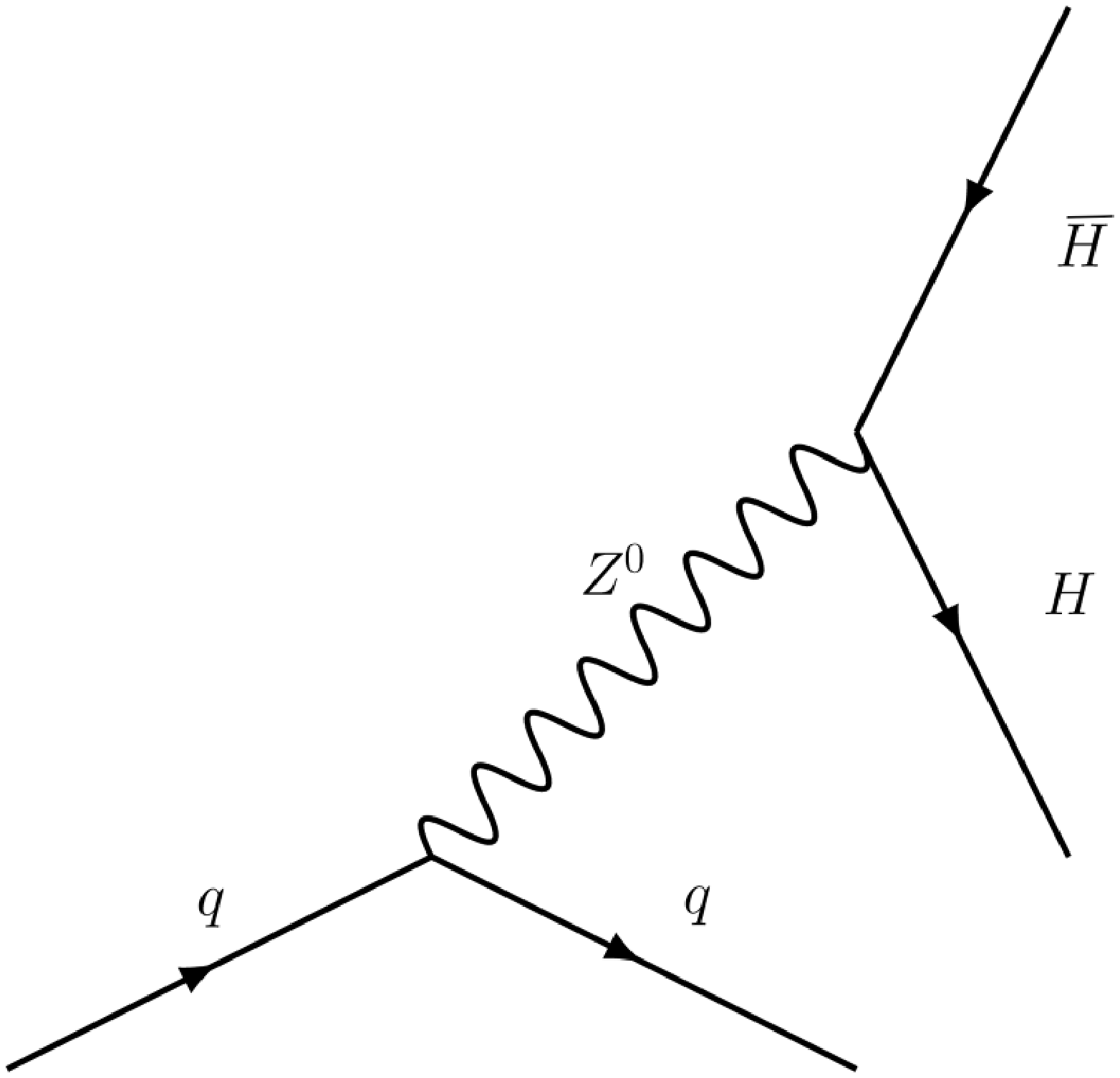} 
\includegraphics[width=0.49\textwidth]{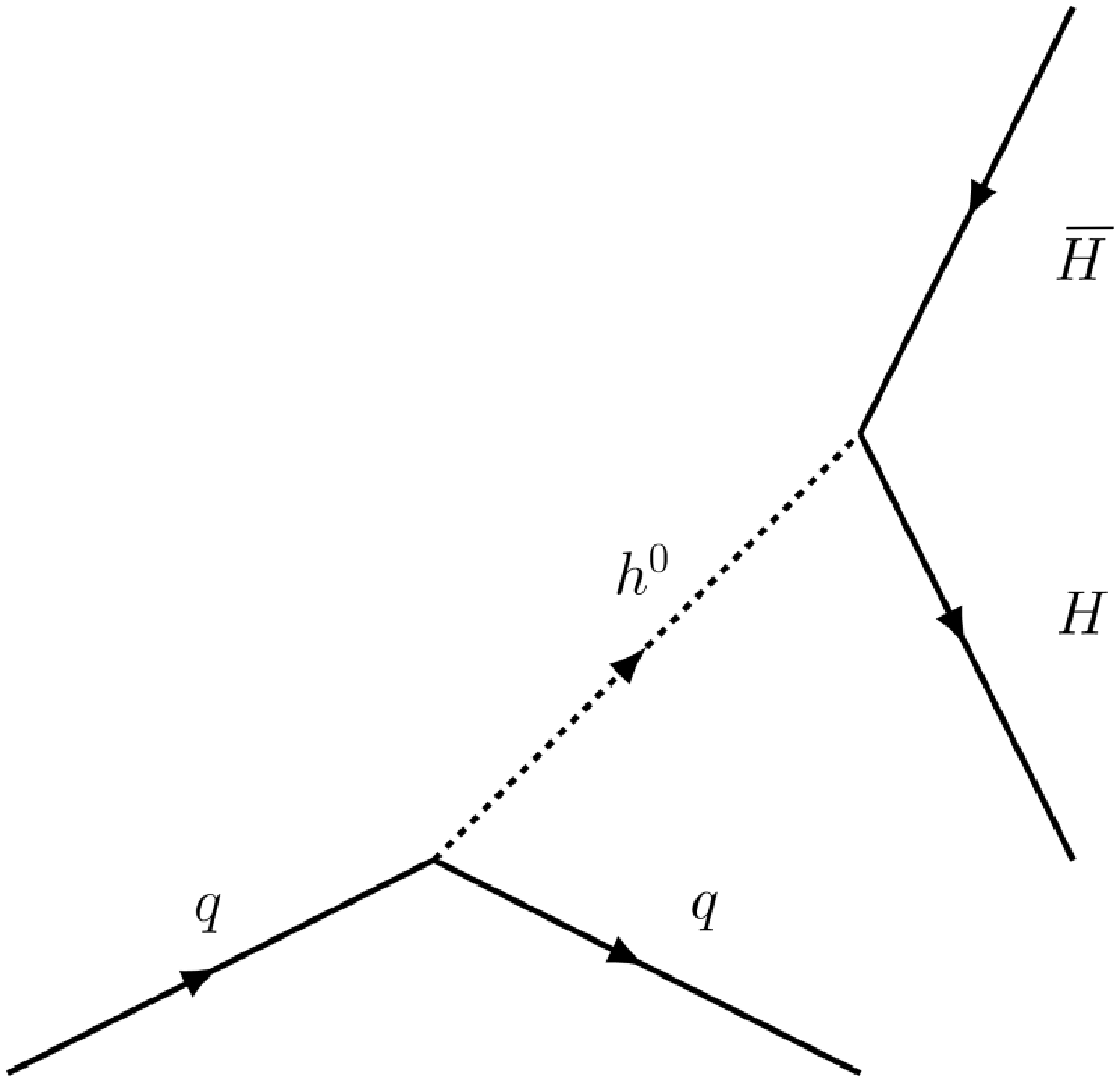} 
\label{fig5}
\caption{Left: Collider production via $Z^0$. Right: Collider production via $h^0$.}
\end{figure}

Here we point out that there should also be an interaction with the recently observed Higgs boson $h^0$. This is consistent with the quartic self-interaction of the Higgs field $\phi$:
\begin{eqnarray}
\mathcal{L}_{\phi}=\lambda_h \left( \phi^{\dag} \left( x \right) \phi \left( x\right) \right) ^2
\label{phi-quartic}
\end{eqnarray}
with
\begin{eqnarray}
\phi =\left( 
\begin{array}{c}
\phi ^{+} \\ 
\phi ^{0 }
\end{array}
\right) \; .
\label{phi}
\end{eqnarray}
In the present theory~\cite{DM1,DM2,statistical},  a scalar field $\phi^r$ represents the amplitude of a 4-component field $\Phi^r$:
\begin{eqnarray}
\Phi^r =\phi^r \, \chi^r   \quad \mathrm{[no \; sum \; on } \; r \mathrm{ ]} 
\label{ampl}
\end{eqnarray}
with
\begin{eqnarray}
\chi^{r \, \dag} \chi ^r=1 \quad \mathrm{[no \; sum \; on } \; r \mathrm{ ]}    \; .
\label{chi}
\end{eqnarray}
Let us focus on the neutral field $\Phi^0$, with condensation of the Higgs field $\phi^0$ plus excitation of a Higgs boson $h^0$ and a Higgson $H^0$. The simplest generalization which yields (\ref{phi-quartic}) (and which has the correct symmetries) is
\begin{eqnarray}
\mathcal{L}_{int} &=& \lambda_h \left( \phi ^{0 \, *} \, \phi ^0 + H^{0 \, \dag } \, H^0  \right) ^2 
\label{H-phi}
\end{eqnarray}
with $\phi^0 = v+ h^0$, where $v$ is the vacuum expectation value of the Higgs field. (Both $\phi^0$ and $H^0$ are dimension 1 bosonic fields.) It follows that there is a lowest-order interaction of the Higgson with the Higgs, given by
\begin{eqnarray}
\mathcal{L}_{H h } &=& 4 \lambda_h v H^{0 \, \dag } \, h^0 \, H^0  \; .
\label{Hh}
\end{eqnarray}
The Higgson $H^0$  then interacts with quarks and other fermions through an exchange of Higgs bosons $h^0$ as well as vector
bosons ($W^{\pm}$ and $Z^0$).

The neutralino $\chi ^{0}$ also has an interaction through the Higgs, if 
$\chi ^{0}$ has an appreciable admixture of both a Higgsino $\widetilde{h}^{0}
$ and a zino $\widetilde{Z}^{0}$, resulting from a term
\begin{equation*}
\mathcal{L}_{\chi h}=\lambda _{\chi }\widetilde{h}^{0\,\dag } \,h^{0}\,\widetilde{Z}^{0}\;.
\end{equation*}
(Both $\widetilde{h}^{0}$ and $\widetilde{Z}^{0}$ are dimension 3/2 fermionic fields.) 

\section{Relative cross-sections for the two varieties of WIMPs}

We can now estimate the ratio of each cross-section for a Higgson $H^{0}$ to
the corresponding one for a neutralino $\chi ^{0}$, by comparing the order
of magnitude of the contributions from external lines and vertices in the
Feynman diagrams of Figs. 1-5 and their neutralino counterparts. 

The external
lines for a $H^{0}$ pair contribute (in order of magnitude) 1, since the normalization for this particle is the same as for a scalar boson. The 
external lines for a $\chi ^{0}$ pair (again in order of magnitude) contribute
the neutralino mass $M_{\chi }$, since these are Majorana fermions. The $H^{0},$\ $Z^{0}$ vertex of Fig. 1 is
momentum-dependent, and therefore makes a contribution that can be
represented as $p_{W}g_{w}$, where $g_{w} \sim 0.1$ is\ a weak coupling constant and 
$p_{W}$ is the WIMP momentum loss, which is of order $10^{-3}M_{H}$ at very best, in natural
units. For $M_{H} \sim\ M_{\chi }$, we conclude that the amplitude
represented by Fig. 1 is typically lower by a factor of $\lesssim 10^{-3}$ compared to its
neutralino counterpart,  and the cross-section is consequently lower by 
$\lesssim 10^{-6}$. The reason for this enormous decrease is that the 
coupling of $H^0$ to $Z^{0}$ is first-order but momentum-dependent.

$H^{0}$ also has second-order couplings (which are
momentum-independent), as reflected in Fig. 2. But these
contributions are even smaller, because they involve two $Z$ or $W$
propagators in addition to two factors of $g_{w}$. An extra factor of 
$g_{w}/M_{Z}$ or $g_{w}/M_{W}$ will reduce 
the cross-section by many orders of magnitude..

Furthermore, these gauge interactions are most relevant for
spin-dependent scattering, which is weak even for the neutralino. The
final conclusion, then, is that the gauge interactions lead to
cross-sections that are hopelessly small for direct detection.

On the other hand, the interactions via the Higgs in Fig. 3 
are comparable to what they are for the neutralino: In
the processes involving $h^{0}$ exchange, there are factors of roughly 1
from the external lines and $\lambda _{H}\,v$ from the vertex. For the
neutralino, the external lines and vertex respectively contribute roughly 
$M_{\chi }$ and $\lambda _{\chi }$. The product is then roughly the same 
if $M_{\chi } \sim v \approx 250$ GeV and $\lambda _{\chi } \sim \lambda _{H} \sim 0.1$.

The quarks making the largest contribution in the left panel of Fig. 3 are those 
which have relatively large masses (and thus relatively large
Yukawa couplings to the Higgs), but which also occur in reasonable
abundance within a nucleon, with the strange quark apparently being optimal. 

The process in the right panel of Fig. 3 takes
advantage of the extremely large mass of the top, with an enormous Yukawa
coupling that compensates for the need to go to higher order, using the
coupling of the gluons $g$ to a nucleus in the detector. 

We conclude that the Higgson $H^{0}$ and the neutralino 
$\chi ^{0}$ should have comparable cross-sections for Higgs exchange. Since this is the dominant process for
direct detection via spin-independent scattering, $H^{0}$ is in the same
basic range of detectability as the neutralino. 

The amplitudes can easily differ by an order of magnitude, however, and the
cross-sections by two orders of magnitude, so quantitative calculations are
needed. The mass of the $H^{0}$ is $\leq 125$ GeV and its coupling constant 
$\lambda _{H}$ is related to the quartic coupling constant of
the Higgs (about 1/6), so better estimates are feasible. 

Fig. 4 shows the simplest processes resulting from the
annihilation of $H^{0}$ with its antiparticle $\overline{H}^{0}$, producing 
a fermion-antifermion pair. Again, for slowly moving dark
matter particles the momentum dependence of the first vertex leads to a tiny
cross-section in the left panel of Fig. 4, but the process in the right panel occurs with
comparable amplitudes for $H^{0}$ and $\chi ^{0}$.

Finally, in Fig. 5, two of a vast number of possible processes are
shown for production at the LHC. The momenta can now be large, and the
momentum-dependent vertex can therefore be comparable to the
corresponding vertex for neutralino production. The present theory also
predicts production of $H^{\pm }$ particles, which will be more readily
detectable but which presumably have much higher masses.

\section{Conclusion}
Supersymmetry predicts a doubling of particles and thus a doubling of the range of physics. The theory of Refs.~\cite{DM1,DM2,statistical} retains this prediction, and also predicts another class of new particles, resulting from an extended Higgs sector. The lowest in mass of these new particles will be stable (with an R-parity of $-1$) if its mass $m_H$  is lower than that of the lowest mass superpartner. This is likely, since $m_H \le m_h$~\cite{DM2}, where $m_h=125$ GeV/c$^2$ is the mass of the observed Higgs boson. With a well-defined mass,  which is in an optimal range for direct detection, and a substantial estimated cross-section via Higgs exchange for many relevant processes, it appears to be ideally suited for direct, indirect, and collider-based detection within the foreseeable future. 

\section*{Acknowledgement} REA benefitted greatly from discussions with Howard Baer and Keith Olive.

\end{document}